\documentclass{article}
\usepackage{graphicx}
\usepackage{amsmath}

\newtheorem{theorem}{Theorem}

\input{tcilatex}

\begin{document}

\title{Particle versus Field Structure in Conformal Quantum Field Theories}
\author{Bert Schroer \\
presently CBPF, Rua Dr. Xavier Sigaud, 22290-180 Rio de Janeiro, Brazil \\
email: schroer@cbpf.br\\
Prof. emeritus of the Institut f\"{u}r Theoretische Physik\\
FU-Berlin, Arnimallee 14, 14195 Berlin, Germany}
\date{May 9, 2000}
\maketitle

\begin{abstract}
I show that a particle structure in conformal field theory is incompatible
with interactions. As a substitute one has particle-like exitations whose
interpolating fields have in addition to their canonical dimension an
anomalous contribution. The spectra of anomalous dimension is given in terms
of the Lorentz invariant quadratic invariant (compact mass operator) of a
conformal generator $R_{\mu }$ with pure discrete spectrum. The perturbative
reading of $R_{0\text{ }}$as a Hamiltonian in its own right i.e. associated
with an action in a functional integral setting naturally leads to the AdS
formulation. The formal service role of AdS in order to access CQFT by a
standard perturbative formalism (without being forced to understand first
massive theories and then taking their scale-invariant limit) vastly
increases the realm of conventionally accessible 4-dim. CQFT beyond those
for which one had to use Lagrangians with supersymmetry in order to have a
vanishing Beta-function.
\end{abstract}

\section{A few introductory remarks}

Ideas about the use of conformal quantum field theory entered particle
physics for the first time at the height of the Kramers-Kronig dispersion
relations \cite{Kastrup}. They were met with reactions ranging from doubts
to outright rejection and the subject lay dormant for another 10 years when
it reemerged on the statistical mechanics side in connection with second
order phase transitions.

In the next section we will show that these early doubts of the old-time
particle physicists were partially justified, because the particle structure
in CQFT is indeed incompatible with interactions. However far from supplying
a coffin nail for its utility in high energy physics, this no-go theorem
also contains the message that one must use finer concepts in order preserve
the usefulness of conformal quantum field theory as a theoretical laboratory
for particle physics. There are massive particle-like objects
(``infraparticles'' \cite{Bu}) which have a continuous mass distribution
with an accumulation of spectral weight at $p^{2}=m^{2}$ whose generating
local fields have an anomalous non-integer (non-semi-integer in the case of
Fermion fields) contribution to their long distance behavior. In a CQFT long
and short distance behavior coalesce and the accumulation of spectral weight
at $p^{2}=0\,$\ which becomes related to the anomalous dimension of
operators is the vestige of the particle interaction in the massive parent
theory from which the CQFT arose by taking the scale-invariant limit. This
structure is the collective effect of a total collapse of all multiparticle
thresholds on top of each other. The standard LSZ large time scattering
limit does not commute with this scaling limit, in fact the LSZ limit of
such fields vanishes. It is believed that in order to re-extract from such a
situation anything which resembles particle physics one has to apply a more
general form of scattering theory \cite{Bu} which is based on expectation
values and probabilities for inclusive cross sections (where outcoming
``stuff'' below a prescribed energy-momentum resolution is not registered)
rather than on amplitudes. But it is presently not clear how one can achieve
this. In the case of infraparticles (the electron in QED which is inexorably
linked to its photon-cloud) where one also meete a situation of coalescing
thresholds, this generalized scattering theory is known to be very useful 
\cite{Bu}.

Recently there has been a quite different and conceptually\footnote{%
The attribute ``conceptually'' here refers to the local quantum physical
aspects and not to differential-geometric ones.} less ambitious but formally
quite attractive idea which promises to strengthen the utility of CQFT for
particle physics and which is presented in the third section. It basically
consists in finding a theory which radically reprocesses the spacetime
interpretation and degrees of freedom of CQFT in such a way that now the
``energy momentum vector'' $R_{\mu }$ of the Dirac-Weyl compactified world $%
\bar{M}$ becomes the bona fide energy momentum instead of $P_{\mu }$ which
in standard canonical or functional terminology means that $R_{\mu }$ is the
one related to an action and not $P_{\mu }$. If one insists that this total
reshuffling of physical interpretation should leave the basic mathematical
building blocks (a certain generating set of algebras and the symmetry group
structure) untouched, then there is only one answer: an associated anti De
Sitter (AdS) theory \cite{Wit}. The nontrivial reprocessing leads to a
mathematical isomorphism as described in \cite{Reh1} i.e. it goes far beyond
that picture about the AdS-CQFT correspondence which is limited to the
(infinitely remote) boundary of AdS (see in particular the remarks at the
end of \cite{Reh2}). The AdS appearance of the AdS structure as a kind of
reprocessed CQFT is less surprizing if one recalls the 6-dimensional
lightcone formalism which one uses in order to obtain an efficient
description of the conformal compactification $\bar{M}$ of Minkowski space $%
M $ and the construction of its covering $\tilde{M}$ \cite{Schroer}.

In this way one obtains a (perturbative) new constructive non-Lagrangian
access to CQFT which opens a new window into the realm of CQFT beyond those
few 4-dimensional Lagrangian candidates for which one had to use a
combination of gauge theory with supersymmetry. This means that one has no
guaranty that the conformal side at all permits a description in terms of an
action.

\section{Particle Structure and Triviality}

We start with recalling an old theorem which clarifies the relation between
the particle-versus-field content of conformal field theories. To be more
precise the following statement is a result of the adaptation of a
combination of several theorems \cite{BF}\cite{Pohl}

\begin{theorem}
The existence of one-particle states in conformally invariant theories
forces the associated interpolating fields to be canonical free fields. The
only particle-like structures consistent with interactions are hidden in the
structure of those interpolating fields which have anomalous dimensions and
whose mass spectrum is continuous with an accumulation of weight at $%
p^{2}=0,\,\,p_{0}>0.$
\end{theorem}

The easiest way to get a first glimpse at this situation is to look at
conformal two-point functions 
\begin{equation}
\left\langle \psi (x)\psi ^{\ast }(y)\right\rangle =\left\{ 
\begin{array}{c}
c\frac{1}{-\left( x-y\right) ^{2}},\,\,dim\psi =1 \\ 
c(\frac{1}{-\left( x-y\right) ^{2}})^{d_{\psi }},\,\,dim\psi =d_{\psi }>1
\end{array}
\right.  \label{an}
\end{equation}
In the first case the application of the LSZ large time scattering limit
yields 
\begin{equation}
\left\langle \psi (x)\psi ^{\ast }(y)\right\rangle =\left\langle \psi
_{in}(x)\psi _{in}^{\ast }(y)\right\rangle
\end{equation}
which preempts the equality $\psi =\psi ^{in}=\psi ^{out},$ whereas in the
anomalous case the large distance fall-off is too strong in order to be
reconcilable with the mass shell structure of a zero mass particle which
means 
\begin{equation}
\psi (x)\overset{LSZ}{\rightarrow }0
\end{equation}
It is worthwhile to reconsider the argument which leads to the absence of
interaction in the space created by the interpolating field $\psi .$ The
crucial observation is that the presence of a zero mass scalar particle
state vector $\left| p\right\rangle $ with 
\begin{equation}
\left\langle p\left| \psi \right| 0\right\rangle \neq 0
\end{equation}
forces $\psi $ to have a two-point function with a canonical scale dimension
dim$\psi =1.$ The special feature of conformal invariance is that this
implies that the two-point function is free i.e. 
\begin{equation}
\left\langle 0\left| \psi ^{\ast }(x)\psi (y)\right| \right\rangle =c\frac{1%
}{\left[ -(x-y-i\varepsilon )\right] ^{2}}
\end{equation}
Such a conclusion relating canonical short distance dimension with absence
of interactions cannot be drawn in the massive case. However the following
theorem which was proven in the late 50$^{ies}$ by Jost and the present
authors, and can be found in \cite{St-Wi}, holds for both cases:

\begin{theorem}
The freeness of the $\psi $ two-point function implies the field $\psi $ to
be a free field in Fock space.
\end{theorem}

The guiding idea is to show that a localized operator or pointlike field
which vanishes on the vacuum, vanishes automatically on all states i.e. is
the zero operator. This is a consequence of the Reeh-Schlieder theorem \cite
{St-Wi} which in conformal field theory is also known under the name
state-field relation). It says that the operators from a region with a
nontrivial causal complement (or fields smeared with test functions with
support in such a region) act cyclically on the vacuum (and on any other
finite energy state). If we denote by $\mathcal{A}(\mathcal{O})$ either the
polynomial $^{\ast }$-algebra of unbounded smeared fields with supports of
testfunctions in $\mathcal{O}$ or the affiliated bounded operator algebra,
this cyclicity property reads 
\begin{equation}
\overline{\mathcal{A}(\mathcal{O})\Omega }=\mathcal{H}
\end{equation}
where the bar denotes the closure and $H$ is the Hilbert space generated by
all fields (bosonic and fermionic). Since (for fermionic $\psi $ there will
be a change of sign) 
\begin{equation}
\psi (x)\mathcal{A}(\mathcal{O})\Omega =\mathcal{A}(\mathcal{O})\psi
(x)\Omega
\end{equation}
if we choose $O$ spacelike with respect to $x,$ the vanishing of the
``current'' $j(x)=(\partial _{\mu }\partial ^{\mu }+m^{2})\psi (x)$ on the
vacuum implies the vanishing on the dense set $\mathcal{A}(\mathcal{O}%
)\Omega $ and hence (operators in physics are closable) on all $\mathcal{H}%
.\,$The next step consists in proving that the commutator of two $\psi s$ on
the vacuum is a c-number 
\begin{equation}
\left( \left[ \psi (x),\psi (y)\right] -i\Delta (x-y)\right) \Omega =0
\end{equation}
It then follows according to the previous argument that the bracket vanishes
identically. We prove this last relation by using the frequency
decomposition $\psi =\psi ^{(-)}+\psi ^{(+)}$ (which follows from $j\equiv
0) $ in the commutator 
\begin{equation}
\left[ \psi (x),\psi (y)\right] \Omega =(\left[ \psi ^{(+)}(x),\psi ^{(+)}(y)%
\right] +\psi ^{(-)}(x),\psi ^{(+)}(y)-\psi ^{(-)}(y),\psi ^{(+)}(x))\Omega
\end{equation}
where we omitted all annihilation terms. The on-shell creation with
subsequent on-shell annihilation as in the last two terms and the physical
spectrum condition only admits the vacuum as its energy momentum content and
therefore they yield a c-number which, by a finite renormalization of $\psi $
if necessary, yields 
\begin{equation}
(\psi ^{(-)}(x),\psi ^{(+)}(y)-\psi ^{(-)}(y),\psi ^{(+)}(x))\Omega =i\Delta
(x-y)\mathbf{1}\Omega
\end{equation}
Since this and the full commutator is causal, the first term on the right
hand side has to vanish all by itself. But on the other hand it is the
separate Fouriertransform of momenta which lie on the forward mass shell and
hence it is the boundary value of an analytic function in two complex
4-vectors of the form $z=\xi -i\eta ,\eta $ from the forward light cone.
However an analytic function which vanish on an open set on its boundary
vanished identically (generalized Schwartz reflection principle). The
resulting relation on the vacuum holds according to the previous arguments
for the operators and therefore we obtained the characterizing relation for
a free field. The generalization to any spin including half-integer values
is now a routine matter. A closer look at the zero mass situation reveals
that contrary to the massive case where the difference of two on-shell
vectors is either spacelike or zero, the difference of two lightlike vectors
may in addition be lightlike but this only happens for parallel vectors.
Since this special configurations should not matter in the sense of L$^{2}$%
-integrability of zero mass particle wave functions one again expects at
least for $d>1+1$ the above result. However a mathematical proof of this
result turned out to be quite nontrivial \cite{Pohl}.

It is very helpful to place the above theorem into the setting of a more
general theorem relating interactions and particle properties in general
local quantum physics which states that operators localized in sub-wedge
regions in interacting theories which possess nontrivial matrix elements
between vacuum and one-particle states necessarily show the phenomenon of
vacuum polarization i.e. operators which create polarization-free
one-particle states exist only in interaction free field theories.
Polarization-free-generators (PFG) which create pure one-particle states
from the vacuum do however exist in any QFT if their localization region is
a semi-infinite wedge region or larger \cite{Essay}\cite{BBS}. Since in
conformal theories the wedge region is conformally equivalent to a compact
double cone, a conformal one-particle structure according to this more
general theorem is only possible in conformal free field theories.

The above argument is typical for a real-time structure which cannot be
unraveled in the euclidean formulation.

\section{Trying to make the best out of it}

The negative result on the compatibility of zero mass particle structure
with nontriviality of conformal theories should not be misread as an
incompatibility with an intuitive idea about what constitutes particle-like
excitations. The point here is that conformal theories in particle physics
should be considered as the zero mass (scaling) limits of massive theories
with mass gaps for which the LSZ scattering theory can be derived. In the
scaling limit all the multiparticle thresholds in momentum space coalesce on
top of each other and build up the possibly anomalous dimension. In this
limit the Wigner particle theory (irreducible representation of the
Poincar\'{e} group) and with it the prerequisite of the LSZ scattering
theory gets lost in the presence of interactions, a fact which we have
demonstrated above where it was shown that the field is either free or the
LSZ limits are zero. So the right question would be: can one think of a more
general scattering theory which may recuperate some of the lost structure in
the aforementioned collapse of multiparticle cuts on top of each other?
There is indeed another particle concept (``infraparticle'') which goes
together with a generalized scattering theory build on inclusive scattering
probabilities instead of amplitudes \cite{Bu}. This concept is expected to
distinguish those anomalous dimensional fields which are of relevance in
particle physics (which originate from the previous collapse in the scaling
limit) from mere mathematical constructs as e.g. generalized free fields
with anomalous dimensions. But we think that for the problem at hand, namely
the formulation of a theory of anomalous dimension, we do not need to enter
this deep and difficult issue of particle-like interpretation since here we
restrict our interests in conformal theories as a simplified theoretical
laboratory for field- and algebra- aspects and not for the study of
particles and their scattering theory. We believe that the setting of local
observable algebras which fulfill in addition to Einstein causality also
Huygens principle for timelike distances \cite{Sch} contains all scale
limits of theories which are of interest for particle physics and that
interaction in this setting is characterized by the appearance of
charge-carrying fields with anomalous dimensions. In view of the above No-Go
theorem we will consider the noncanonical (anomalous dimension) nature of
those fields as our pragmatic definition of interaction in this conformal
setting. But we defer this analysis to a following longer paper which
contains the relevant mathematical machinery \cite{Sch}.

As a consequence the observable algebra of an interacting conformal field
theory (conserved currents etc.) should not have the structure of composites
of free fields (e.g. free currents) since otherwise the fields carrying the
superselected charges may not have anomalous dimensions. Apart from
normalization constants the 2- and 3-point functions of conformal observable
fields (currents) are indistinguishable from those formed with free
composites with the same integer dimensions. If all correlations would be
indistinguishable from those of free composites (total protection) then also
the charge-carrying fields associated with such observables can be shown to
be free.

A weak form of what in the case of conformal SYM theories has been called
(partial) ``protection'' would be one where the relative normalization
between 2-and 3-point functions is that of free composites (partial
protection). Apparently perturbative supersymmetry causes partial
protections \cite{prot}. Although such models hardly represent realistic
particle physics, they are the only \textit{Lagrangian} candidates for d=1+3
nontrivial conformal field theories and may yet turn out to be the first
4-dimensional mathematically completely controllable models. The interest
and fascination in conformal field theories originates to a large part from
the well-founded belief that the simplest nontrivial 4-dimensional conformal
field theories which will break the age old existence deadlock\footnote{%
In any area of Theoretical Physics there always have been plenty of
nontrivial mathematically controllable illustrations which demonstrate the
nontrivial physical content of the conceptual basis of those areas, not so
in 4-dim. QFT. This annoying totally singular situation has been sometimes
overemphasized at the cost of practical calculations, but most of the time
it went totally ignored.} for nontrivial quantum field theories in physical
spacetime. For this one wants to have as much protection as possible without
ending with a free conformal theory.

Instead of entering an ambitious program in order to extract the particle
physics ``honey'' from CQFT which requires a heavy conceptual investment in
the area of a generalized scattering theory, there is another way which is
more faithful to the formal aspects with which QFT is often identified
(erroneously in my opinion, if one uses them for a definition of QFT) namely
canonical formalism and/or functional integrals. It starts from the
observation that in addition to the translation generator $P_{\mu }$ there
is another translation-analogue described by a Lorentz-vector $R_{\mu }.$ It
has a timelike purely discrete spectrum and the L-invariant ``mass'' $m_{c}$
with $m_{c}^{2}=R_{\mu }R^{\mu }$ plays a similar role as the rigid rotation
operator $L_{0}$ in chiral theories. In fact it describes a generalized
rotation around the Dirac-Weyl compactified Minkowski space $\bar{M}\simeq
S^{3}\times S^{1}.$ Therefore it is not surprising that the bottom of the
spectrum of $m_{c}$ is the anomalous part of the scaling dimension common to
a whole equivalence class of fields which carry the same superselected
charge. But despite all analogies to $P_{\mu }$ this operator is not related
to an imagined functional integral action of CQFT. Nevertheless one can ask
the question: is there a theory whose Lagrangian can be associated with a
Hamiltonian interpretation of $R_{0}?$ In order for this new theory to be
useful for particle physics it should keep the same algebraic and
group-theoretical building blocks as CQFT i.e. one seeks a mathematical
isomorphism which goes hand in hand with that total physical reprocessing
which is necessary to accomplish such an impossible looking task. The unique
answer is the AdS-CQFT correspondence \cite{Wit} which was proven to be a
such a ``radical'' isomorphism \cite{Reh1}.

Although this step does not completely answer the question posed at the
beginning of how to extract and analyze the particle content of CQFT, it
goes a long way to open up conformal field theory as a genuine theoretical
laboratory for particle physics. And last not least it facilitates the
unsolved problem number one: find a nontrivial physically relevant (i.e. one
which fits at least the conceptual framework of local quantum physics, even
if it falls short in describing nature) and mathematically controllable
model in 4-dimensional QFT.

The presented arguments suggest strongly that there exists a whole world of
non-Lagrangian non-supersymmetric CQFT (in the sense that they cannot be
accessed in the standard perturbative way) besides the Lagrangian SYM
family. In fact the perturbative calculations in the literature already give
some support in this direction. This is most visible in \cite{Ruehl}
although these authors, evidently under the strong spell of the
string-theoretic origin of the AdS-CQFT, do not interprete their
calculations from this viewpoint.

The possible non-Lagrangian nature of most CQFT is in a certain way
explained by Rehren's deep observation \cite{Reh1}\cite{Reh2} that due to
the isomorphic nature of the AdS-CQFT relation there must be degrees of
freedom on the conformal side which cannot be described in terms of local
fields namely those which originate from the AdS bulk (and not from the
boundary) and which are necessary in order to return $CQFT\rightarrow AdS$.
This leaves the interesting question of what should one make of the original
observation by which the protagonists of the AdS-CQFT correspondence found
this relation which is the relation between two Lagrangian field theories
namely the conformal SYM model with some form of AdS supergravity \cite{Wit}%
. Since this is based on consistency checks within string theory which owes
its widespread acceptance to perturbative mathematical consistency and a
kind of globalized social contract but certainly not to its harmonious
coexistence with the principles underlying particle physics, there is reason
for some scepticism; in particular because such degrees of freedom would be
easily overlooked in perturbative calculations on the CQFT side. It cannot
be overstressed that this correspondence is very different and much more
radical then those which arise from a different choice of \ ``field
coordinates''. It is impossible to understand its full content in terms of
pointlike physical fields.  

\section{Some concluding remarks}

If, as argued in this letter, the AdS theories are a useful new
calculational tool which open up CQFT to particle physics studies within the
standard Lagrangian quantization framework, than perhaps with an additional
conceptual investment one could directly understand the structure underlying
the anomalous dimension spectra within CQFT i.e. without the described
reprocessing on the AdS side. This turns out to be true and will be the
subject of a subsequent paper \cite{Sch} since the necessary conceptual
investment does not fit the format of a letter like this.

\textit{Acknowledgements}: I am indebted to Detlev Buchholz and Karl-Henning
Rehren for a helpful exchange of emails. Furthermore I would like to thank
Francesco Toppan for interesting questions which helped in shaping the
presentation.

\end{document}